\documentstyle[epsfig]{article}

\begin{document}
\begin{center}
    {\large\bf $M$-body Pure State Entanglement}
\vskip .6cm {\normalsize Feng Pan,$^{a,b,}$ and J. P. Draayer$^{b}$
} \vskip .2cm {\small $^{a}$Department of Physics, Liaoning Normal
University, Dalian 116029, P. R. China\vskip .1cm $^{b}$Department
of Physics and Astronomy, Louisiana State University, Baton Rouge,
LA 70803-4001}
\end{center}
\vskip .5cm {\bf Abstract:~}\normalsize The simple entanglement of
$N$-body $N$-particle pure states is extended to the more general
$M$-body or $M$-body $N$-particle states where $N\neq M$. Some new
features of the $M$-body $N$-particle pure states are discussed. An
application of the measure to quantify quantum correlations in a
Bose-Einstien condensate model is demonstrated.

\vskip .3cm \noindent {\bf Keywords:} Many-body pure state,
entanglement measure, quantum correlations. \vskip 0.3cm \noindent
{\bf PACS numbers:} 03.67.-a; 03.67.Mn; 03.65.Ud; 03.65.Bz \vskip
.5cm

\section{Introduction}

Entanglement plays an important role in quantum information theory
and quantum computation~\cite{[1]}-\cite{[3]}. Entanglement normally
considered is about $N$-body $N$-particle states. The most famous
and widely studied of these are probably the Bell states~\cite{[4]}
with $N=2$, and $N$-body $N$-particle GHZ~\cite{[5]} and
W~\cite{[6]} states with $N\geq 3$. Often, however, that the number
of particles $N$ is not equal to the number of kinds of particles
$M$ in a quantum state. Such multipartite entangled states can be
generated naturally in numerous many-body quantum systems, such as
spin systems~\cite{[7]}-\cite{[9]}, one dimensional chains or
lattices of various dimensions of Bose or Fermi many-body
systems~\cite{[10]}-\cite{[12]} including two-body $N$-particle
entangled states generated in the Bose-Einstein
condensate~\cite{[13]} with sufficiently large $N$, of which
physical systems can be fabricated with modern technology.
Specifically, entangled states can be realized by burning the system
into a desired configuration and then cooling it down, which leads
to the so-called ground state entanglement~\cite{[14]}. In order to
classify different entangled states and study the entanglement of
these types of states, it is necessary to extend the definition of
entanglement to $M$-body $N$-particle pure states where $N\neq M$.
Up till now, there have been at least four different approaches to
the problem, namely, the entanglement of modes~\cite{[15]}, the
quantum correlations~\cite{[16]}, the entanglement of
particles~\cite{[17]}, and entanglement witness~\cite{[18]}.  The
entanglement of modes proposed by Zanardi~\cite{[15]} is defined in
terms of local entanglement in terms of local reduced density matrix
in Fock representation. The quantum correlations for pure states
introduced in [16] are determined by the so-called Slater rank of
the state. The entanglement of particles provides an operational
definition of entanglement between two parties who share an
arbitrary pure state with a standard measure~\cite{[17]}. Finally,
entanglement of witness is defined in [18] based on witness
operators. Each of these approaches emphasizes one or several
respects of the problem. There is still no consensus reached on
which one is the most suitable.

As an extension of the entanglement measure for $N$-body
$N$-particle pure states~\cite{[19]}, in this paper, we propose an
entanglement measure for $M$-body $N$-particle states for any $M$
and $N$, which can be regarded as an extension of the entanglement
of modes proposed by Zanardi~\cite{[15]}, but does not count in
entanglement among local identical (indistinguishable) particles. In
Sec. 2, we will propose the entanglement measure for $M$-body
$N$-particle pure states for any $M$ and $N$. Some simple examples
taken from quantum many-body systems will be analyzed with this
measure in Sec. 3. Sec. 4 is a short discussion.

\section{The entanglement measure for $M$-body $N$-particle pure
states}

For a system consisting of $M$ kinds of particles, any pure state
$|\Psi\rangle$ can be expanded in terms of basis vectors
$|n_{1},\sigma_{1};n_{2},\sigma_2;\cdots; n_{M},\sigma_M\rangle$ in
the tensor product space $V_{k_{1}}\otimes
V_{k_{2}}\otimes\cdots\otimes V_{k_{M}}$ as

$$|\Psi\rangle=\sum_{n_{1}\sigma_{1}n_{2}\sigma_{2}\cdots n_{M}\sigma_{M}}
C^{n_{1}n_{2}\cdots n_{M}}_{\sigma_{1}\sigma_{2}\cdots\sigma_{M}}
|n_{1},\sigma_{1};n_{2},\sigma_2;\cdots;
n_{M},\sigma_M\rangle,\eqno(1)$$

\noindent where $n_{i}$ ($1\leq i\leq M$) is the
number of the $i$-th kind of particle, $\sigma_i = 0,1,\cdots,
k_{i}$~ for $1\leq i\leq M$ denotes the internal degrees of freedom, and
$C^{n_{1}n_{2}\cdots n_{M}}_{\sigma_{1}\sigma_{2}\cdots\sigma_{M}}$
is the normalized expansion coefficient. Then (1) will be called an
$M$-body pure state. In addition, if the total number of particles
$\hat{N}=\sum_{i=1}^{M} \hat{n}_{i}$ is a conserved quantity with

$$\hat{N}\vert\Psi\rangle=N\vert\Psi\rangle,\eqno(2)$$
(1) will be called an $M$-body $N$-particle pure state, which will
be denoted as $\vert\Psi(M,N)\rangle$ in the following. According to
the above definition, Bell states are two-body two-particle states,
GHZ and W states are $N$-body $N$-particle states with $N\geq 3$.
The corresponding density matrix of (1) is

$${\rho}_{\Psi}=\vert\Psi\rangle\langle\Psi\vert.\eqno(3)$$

\noindent Let $a_{i\sigma}^{\dagger}$ ($a_{i\sigma}$) be local
particle creation (annihilation) operators that satisfy

$$[a_{i\sigma},a_{j\sigma^{\prime}}^{\dagger}]_{\pm}\equiv
a_{i\sigma}a_{j\sigma^{\prime}}^{\dagger}\pm
a_{j\sigma^{\prime}}^{\dagger}a_{i\sigma}=
\delta_{ij}\delta_{\sigma\sigma^{\prime}},\eqno(4)$$

$$[a_{i\sigma},a_{j\sigma^{\prime}}]_{\pm}=
[a^{\dagger}_{i\sigma},a_{j\sigma^{\prime}}^{\dagger}]_{\pm}=0,
\eqno(5)$$ for fermions or bosons. The wavefunction
$\vert\Psi(M,N)\rangle$ can be expressed as

$$\vert\Psi(M,N)\rangle=\sum_{n_{1}\sigma_{1}\cdots n_{M}\sigma_{M}}
C_{n_{1}\sigma_{1}\cdots
n_{M}\sigma_{M}}(a_{1}^{\dagger})_{\sigma_{1}}^{n_{1}}
(a_{2}^{\dagger})_{\sigma_{2}}^{n_{2}}\cdots
(a_{M}^{\dagger})_{\sigma_{M}}^{n_{M}} \vert 0\rangle,\eqno(6)$$
where $\vert 0\rangle$ is the vacuum state and
$(a_{i}^{\dagger})_{\sigma_{i}}^{n_{i}}\equiv
\sum_{\sigma^{i}_{1}\sigma^{i}_{2}\cdots
\sigma^{i}_{n_{i}}}\beta^{\sigma_{i}}_{\sigma^{i}_{1}\sigma^{i}_{2}\cdots
\sigma^{i}_{n_{i}}}a_{i\sigma^{i}_{1}}^{\dagger}a_{i\sigma^{i}_{2}}^{\dagger}\cdots
a_{i\sigma^{i}_{n_{i}}}^{\dagger}$, in which
$\beta^{\sigma_{i}}_{\sigma^{i}_{1}\sigma^{i}_{2}\cdots
\sigma^{i}_{n_{i}}}$ is the expansion coefficient.

In (1) and (6), $\{\vert n_{i}\sigma_{i}\rangle\}$ is recognized as
a set of the local basis vectors for the $i$-th kind of particles,
which can be used, for example, to describe a set of local states on
the $i$-th site of a lattice in many lattice models. In such cases,
only entanglement among particles on different sites is of interest,
which can also be used to signify quantum correlations in a model
system. Therefore, entanglement among particles on the same site
will not be considered. Similar problems were also considered in
many works. For example, hierarchic classification for arbitrary
multi-qubit mixed states based on the separability properties of
certain partitions was considered by D\"{u}r and Cirac in [20].

Under the replacement $a_{i\sigma^{i}_{j}}^{\dagger}\rightarrow
X_{i\sigma^{i}_{j}}$, where $X_{i\sigma^{i}_{j}}$ is simply a
symbol, the operator form in front of the vacuum state on the
right-hand-side of (6) becomes a homogeneous polynomial of degree
$N$ in terms of the $\{{\bf X}_{i}\}$,

$$F_{C}({\bf X}_{1},\cdots,{\bf X}_{M})
=\sum_{\sigma_{1}\cdots\sigma_{N}} C_{\sigma_{1}\cdots
\sigma_{N}}(X_{1})^{n_{1}}_{\sigma_{1}} \cdots
(X_{M})^{n_{M}}_{\sigma_{M}}.\eqno(7)$$ It should be understood that
${\bf X}_{i}$ is a multi-value symbol with ${\bf
X}_{i}=X_{i\sigma^{i}_{j}}$ for any $\sigma^{i}_{j}$. An alternative
definition of entangled states with respect to the local bases can
be stated as follows: The state $\vert\Psi(M,N)\rangle$ is an
$M$-body $N$-partite entangled state if the corresponding polynomial
$F_{C}({\bf X}_{1},\cdots,{\bf X}_{M})$ on complex field ${\bf\cal
C}$ cannot be factorized into the following form

$$F_{C}({\bf X}_{1},\cdots,{\bf X}_{M})
=F_{A}({\bf X}_{i_{1}},\cdots,{\bf X}_{i_{l}}) F_{B}({\bf
X}_{i_{l+1}},\cdots,{\bf X}_{i_{M}})\eqno(8)$$ for $1\leq l\leq
M-1$, where $\{i_{1}\neq i_{2}\neq\cdots\neq i_{M}\}$ can be in any
ordering of $\{1,2,\cdots,M\}$. Otherwise the state
$\vert\Psi(M,N)\rangle$ is not an $M$-body $N$-particle entangled
state with respect to the $M$ sets of local bases. The state
$\vert\Psi(M,N)\rangle$ given in (6) is disentangled (separable) if
the polynomial $F_{C}$ can be factorized into a product of
polynomials of ${\bf X}_{i}$ as $\prod_{i=1}^{M}F_{A_{i}}({\bf
X}_{i})$. In other cases, the state is partially entangled.

As is shown in [19], a criterion for distinguishing whether a
homogeneous polynomial is factorizable can be established by using
the von Neumann entropy of the reduced local density matrices. As an
extension of the entanglement measure \cite{[19]} for $N$-body
$N$-partical qubit system, we define entanglement measure for the
$M$-body pure state (1) as

$$\eta^{(M)}=\left\{\matrix{
{1\over{M}}\sum^{M}_{i}S_{(i)},~~~{\rm if}~~S_{i}\neq
0~~\forall~~i,\cr \cr~~~0~~~~~~~~~~~~~{\rm
otherwise,}\cr}\right.\eqno(9)$$

\noindent where $M$ is the number of
different kinds of particles involved in (1),

$$S_{(i)}=-{\rm Tr}[\rho_{(i)}\log_{p_{i}}(\rho_{(i)})]\eqno(10)$$ is the
extended local reduced von Neumann entropy expressed in terms of
local reduced density matrix $\rho_{(i)}$ for the $i$-th kind of
particles only obtained by taking the partial trace over the
subsystem, and $p_{i}$ is the total number of the Fock states of the
$i$-th kind of particles in the system. We use the logarithm to the
base $p_{i}$ instead of base $2$ used in qubit system~\cite{[19]} to
ensure that the maximal measure is normalized to $1$. Hence, the
maximal value of the measure is $1$ regardless how many particles
and how many kinds of particles are involved in the system. While
the measure is zero when a pure state is not a genuine $M$-body
entangled state.

 Similar to the
$N$-body $N$-particle case~\cite{[19]}, the measure (9) quantifies
genuine $M$-body entanglement. The measure is zero if $S_{i}=0$ for
any $1\leq i\leq M$, in which the corresponding state is not a
genuine $M$-body entangled one. In contrast to the measure defined
in terms of local von Neumann entropy~\cite{[16]}, which provides
information of local entanglement only, the measure (9) provides
information about overall quantum correlations among $M$ kinds of
particles in a quantum many-body system.

In the pure state (6), the local particle number is not a conserved
quantity in general. In such a case, the local reduced density
matrix $\rho_{(i)}$ is built in the local Fock space spanned by $\{
\vert n_{i}\sigma_{i}\rangle\}$ ($0\leq n_{i}\leq p_{i}$) with the
following block-diagonal form:

$$\rho_{(i)}=\left(\matrix{(n_{i}=0)\cr &(n_{i}=1)\cr
&&(n_{i}=2)\cr&&~~~~~~~~\ddots\cr&&&(n_{i}=p_{i})\cr}\right),\eqno(11)$$
where each sub-matrix $(n_{i}=q)$ is spanned by the $q$-particle
Fock states with the internal degrees of freedoms
$0\leq\sigma_{i}\leq k_{i}$. Therefore, the extended reduced von
Neumann entropy $S_{(i)}$ is not only invariant under the Local
Unitary (LU) transformations for each block in (11) with respect to
internal degrees of freedoms $\{\sigma_{i}\}$ for fixed $n_{i}$, but
also invariant under unitary transformations for the subspace
spanned by the entire Fock states with the local particle numbers
$0\leq n_{i}\leq p_{i}$. Hence, the measure (9) is also invariant
under both the Local Unitary (LU) transformations with respect to
internal degrees of freedoms and unitary transformations for the
subspace spanned by the entire Fock states. In addition, the measure
does not count in entanglement among local identical particles with
the same label $i$ and different internal degrees of freedom because
the local basis vectors are recognized to be $\{\vert
n_{i}\sigma_{i}\rangle\}$.

In the $M=N$ qubit cases discussed in [19], the LU invariance is
equivalent to the invariance under Local Operations assisted by
Classical Communications (LOCCs). In such case the measure (9) is
both LU and LOCC invariant~\cite{[21]}. In $N\neq M$ cases, however,
LOCCs may be useless in any quantum information processing protocol.
For example,  LOCCs will not only keep the entanglement the same,
but also be irrelevant to any quantum information processing
processes for $M\neq N$. In addition, though (9) is invariant under
local unitary transformations with respect to different number of
particles, such transformations will result in local state mixing
among different number of particles, which obviously violates the
super-selection rule, and thus is impossible~\cite{[22],[23],[24]}.
However, the measure (9) is indeed useful in quantifying the
$M$-body $N$-particle entanglement, of which some examples will be
shown in the next section.

\section{Some simple examples}

Using the above definition, one can easily write out many
non-trivial states with $M\neq N$ which are quite different from
those with $M=N$. For simplicity, let us assume that there are $M$
kinds of spinless fermions or bosons. Then, the simplest case is a
$M$-body one-particle state with

$$\vert\Psi(M,1)\rangle=C_{1}\vert \overbrace{10\cdots 0}^{M}\rangle
+C_{2}\vert \overbrace{010\cdots 0}^{M}\rangle+\cdots+C_{M}\vert
\overbrace{0\cdots 01}^{M}\rangle,\eqno(12)$$ where $\vert
{0\cdots010\cdots 0}\rangle$ with $1$ in the $i$-th position denotes
one-particle state for the $i$-th kind of particle and vacuum state
of the remaining particles, and $C_{i}$ ($i=1,2,\cdots,M$) are
normalized non-zero expansion coefficients. These types of states
emerge naturally from many models, such the Bose- and Fermi-Hubbard
models.

In $N$-body $N$-particle cases, generally a state is entangled with
respect to its intrinsic degrees of freedom, such as the third
component of spin of photons or atoms. In such cases, the measure
(9) is invariant under LOCCs or LUs. In $N\neq M$ cases, however,
LOCCs may be useless in any quantum information processing protocol.
In order to make this point clear, for example, let us replace
spinless fermions or bosons in (12) by those with internal degrees
of freedom $\sigma_{i}=1,2,\cdots,\mu_{i}$ for $i=1,2,\cdots,M$.
Then, (12) becomes

$$\vert\Psi(M,1)(\sigma_{1}\sigma_{2}\cdots\sigma_{M})
\rangle=C_{1}\vert \overbrace{1_{\sigma_{1}}0\cdots 0}^{M}\rangle
+C_{2}\vert \overbrace{01_{\sigma_{2}}0\cdots
0}^{M}\rangle+\cdots+C_{M}\vert \overbrace{0\cdots
01_{\sigma_{M}}}^{M}\rangle.\eqno(13)$$ $\mu_{i}$ should be all the
same for any $i$ if these are identical particles, otherwise they
are not identical. In this case, any LOCC between $i$-th and $j$-th
parties is equivalent to a corresponding LU transformation with
respect to the internal degrees of freedom labeled by $\sigma_{i}$
or $\sigma_{j}$. After any such LU transformation, say at $i$-th
party, the vacuum state of the $i$-th particle for any $i$ remains
invariant, while local one-particle state still remains to be
one-particle state with a linear combination of different internal
labels. The $j$-th party can not get any information after local
operation at $i$-th party assisted by classical communication
between them and then measuring the resultant $j$-th local state
because the $j$-th state remains to be the same after any local
operation at $i$-th party. Therefore,  LOCCs will not only keep the
entanglement the same for the state shown in (12), but also be
irrelevant to any quantum information processing processes used in
$M=N$ cases even if those $M$ Fock states have additional intrinsic
degrees of freedom as shown in (13). Actually, $N\neq M$ entangled
states, such as that shown in (13), are entangled with respect to
local particle numbers in $M$ different local Fock states and not
with respect to local intrinsic degrees of freedom. For a state,
such as that shown in (13), effective local operations similar to
those used in $M=N$ cases are local particle number non-conserved
unitary operations, namely local projective operations, $a\vert
0\rangle+b\vert 1\rangle$, where $\vert 0\rangle$ stands for vacuum
state, $\vert 1\rangle$ stands for one particle state, and $a,~b\in
\bf C$ with $\vert a\vert^2+\vert b\vert^2=1$. As for entanglement
in spin degrees of freedom, in which any local operation may violate
conservation of spin in the third component, any local particle
number non-conserved unitary operation violates both local and total
particle number conservation. Such operations obviously violate the
super-selection rule, and thus are impossible to be implemented.
While the entanglement measure (9) is invariant under any such local
particle number non-conserved operation.

Although  local particle number non-conserved operations are
physically impossible, measure (9) can be used to quantify
entanglement of any $M$-body entangled state. For example, according
to (9), the entanglement measure for (12) is
$$\eta^{(M,N=1)}=-{1\over{M}}\sum^{M}_{i=1}\left( \vert
C_{i}\vert^{2}\log_{2}(\vert C_{i}\vert^{2})+(1-\vert
C_{i}\vert^{2})\log_{2}(1-\vert C_{i}\vert^{2})\right).\eqno(14)$$
It can be verified by conditional maximization that (12) reaches its
maximal value when $\vert C_{i}\vert=1/\sqrt{M}$ with $M\geq 2$.
Clearly, in the cases with maximal measure

$$\eta^{(M,N=1)}_{\max}=\log_{2}(M)-{M-1\over{M}}\log_{2}(M-1)
~~{\rm for}~~~M\geq 2.\eqno(15)$$  It is similar to the Bell states
with $\eta^{(2,1)}_{\max}=1$ when $M=2$, and is equivalent to the W
states with $\eta^{(3,1)}_{\max}=0.918296$ when $M= 3$, which is the
same value as that of a W state for a qubit system calculated in
\cite{[25]}.

Though there is controversy about whether such states as that shown
in (12) are entangled or not~\cite{[26],[27]}, it was clearly
demonstrated in \cite{[28]} that such state is indeed a
single-particle entangled state, which was also used in a two-party
protocol for quantum gambling~\cite{[29]}, and is proved to be
useful in the so-called data hiding protocols~\cite{[24]}. The
measure (9) is in agreement with those observations.

Another example of our application of the measure (9) is to study
quantum phase transition in a Bose-Einstien condensate model with
$n$ bosons in a two-dimensional harmonic trap interacting via a
fairly universal contact interaction~\cite{[30],[31], [32]}, of
which the Hamiltonian can be written as

$$H=\sum^{\infty}_{j=0}\omega j a^{\dagger}_{j}a_{j}+
g\sum_{i,j,k,l}\delta_{i+j,k+l}V_{i,j;k,l}a^{\dagger}_{i}a^{\dagger}_{j}
a_{k}a_{l},\eqno(16)$$ where $a^{\dagger}_{i}$ ($a_{i}$) is creation
(annihilation) operator of the $i$-th boson, and
$V_{ij;kl}=(k+l)!/(i!j!k!l!4^{k+l})^{1\over{2}}$. In this model, the
total number of bosons $\hat{n}$ and the angular momentum $\hat{L}$
with

$$\hat{n}=\sum_{i=0}^{\infty}a_{i}^{\dagger}a_{i},
~~\hat{L}=\sum_{j=0}^{\infty}ja_{j}^{\dagger}a_{j}\eqno(17)$$  are
two conserved quantities. Therefore, for given $n$ and $L$,
eigenstates of (16) can be written as
$$
\vert n,L; \tau\rangle=\sum_{n_{2}n_{3}\cdots n_{L}}
c^{\tau}_{n_{2}n_{3}\cdots n_{L}} {a_{0}^{\dagger}}^{
n-L+\sum_{\mu=2}^{L}(\mu-1)n_{\mu}}\times$$ $${a_{1}^{\dagger}}^{
L-\sum_{\mu=2}^{L}\mu n_{\mu}}a_{2}^{\dagger n_{2}} a_{3}^{\dagger
n_{3}}\cdots a_{L}^{\dagger n_{L}}\vert 0\rangle,\eqno(18)$$ where
$c^{\tau}_{n_{2}n_{3}\cdots n_{L}}$ are expansion coefficients, and
$\tau$ is an additional quantum number needed in distinguish
different eigenstates with the same $n$ and $L$.

To explore transitional patterns in the system, we calculate the
entanglement according to (9) along the yrast line up to $L=6$, in
which the local reduced  density matrices $\rho_{(i)}$  for the
$i$-th kind of bosons are obtained by taking the partial trace over
the subsystem. Fig. 1 shows the entanglement measure (9) of the
system as a function of the total number of bosons $n$ along the
yrast line with $L=2,~3,\cdots,~6$. Generally speaking, the maximal
value of the measure decreases with $L$; and the measure for any $L$
decreases with increasing of $n$ due to the $s$-boson dominance in
the yrast states when $n$ is sufficiently large.  It is obvious that
there is always a peak in the measure at or near $n=L+1$ when $L\geq
4$. The peak in the entanglement measure signify that there is a
quantum phase transition occurring near $n=L+1$, in which the total
number of bosons $n$ serves as the control parameter.

In order to show that the peak point in the entanglement measure
shown in Fig. 1 indeed is a critical point of the system, we study
yrast state occupation probabilities $\xi_{l}(n)$ for bosons with
different angular momentum $l$ as a function of $n$ according to

$$\phi_{l}(n)=\langle a^{\dagger}_{l}a_{l}\rangle/n.\eqno(19)$$
The results for $L=2$, $3$, $4$, $5$ cases are shown in Fig. 2,
which shows that the values of the occupation probabilities for
bosons with different angular momentum vary drastically with
increasing of the total number of bosons, especially for $s$- and
$p$-bosons. When $n=L$, the $l$-boson, especially the $p$-boson
components contribute significantly to the yrast states. There are
cross points for yrast state occupation probabilities $\phi_{l}(n)$
with different $l$, which are indeed all near $n=L+1$ point. The
$L=4$ and $L=5$ cases shown in Fig. 2 are consistent to the critical
points at the entanglement measures shown in Fig. 1 though it is not
shown by the entanglement measures in Fig. 1 for $L=2$ and $3$
cases. The results show that there are two different phases. One is
the $p$-boson dominant phase, while another is the $s$-boson
dominant phase, which are controlled by the total number of bosons
in the system. Therefore, the peak in the entanglement measure
indeed signifies that there is a quantum phase transition occurring
near $n=L+1$, in which the total number of bosons $n$ serves as the
control parameter. The result also shows that the measure (9) is
indeed suitable to benchmark the quantum correlations among bosons
with different angular momenta.

\begin{center}
\begin{figure}
\includegraphics[width=9.5cm]{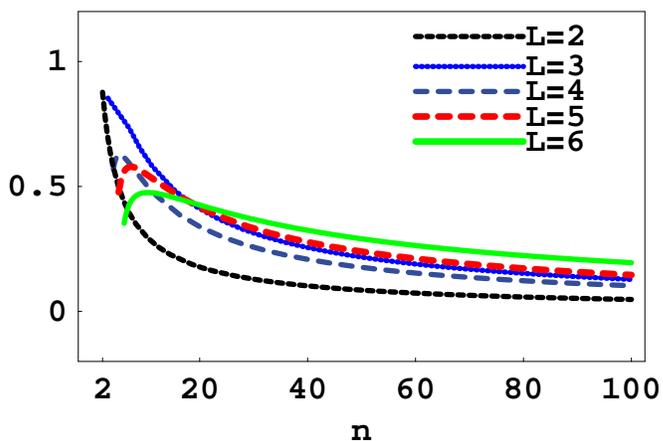}
\caption{ The entanglement measure $\eta^{(L+1)}$ as functions of
the total number of bosons $n$ along the yrast line up to $L=6$.}
\end{figure}
\end{center}
\begin{figure}
\includegraphics[width=8cm]{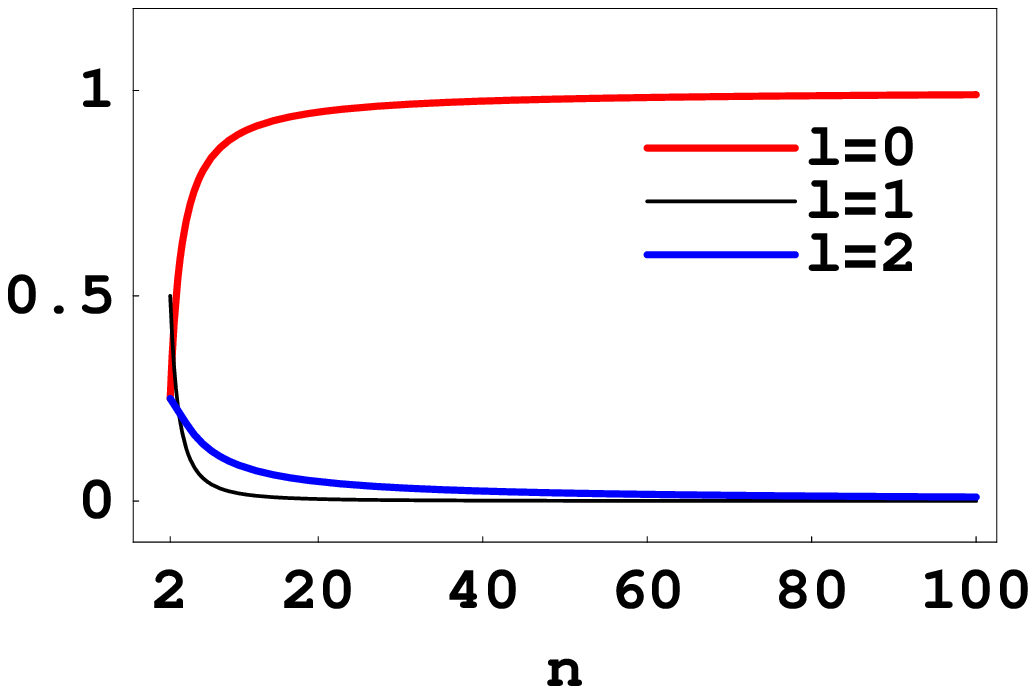}\includegraphics[width=8cm]{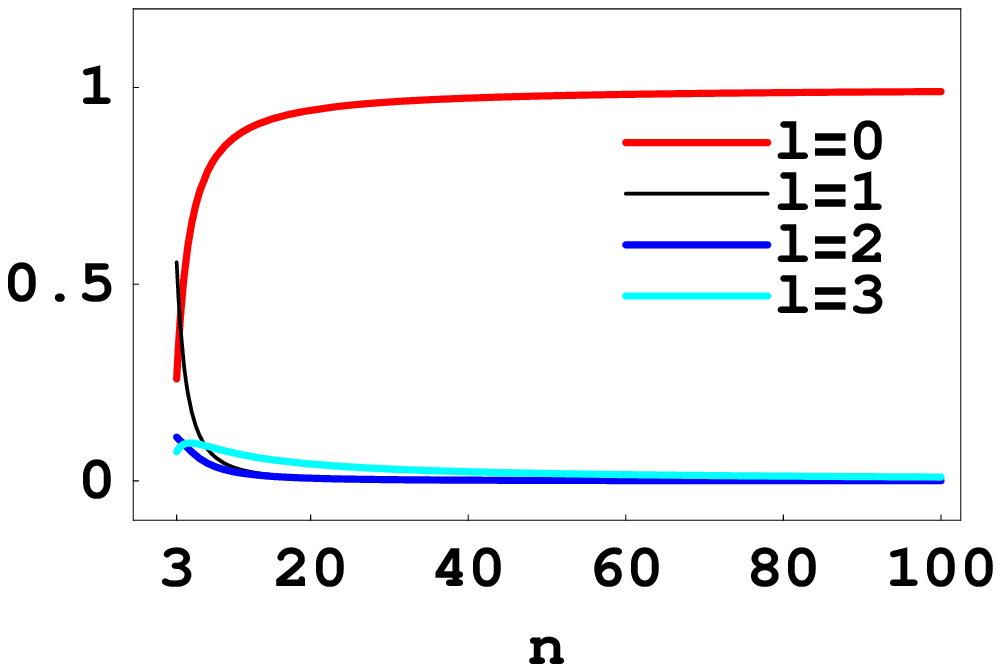}
\includegraphics[width=8cm]{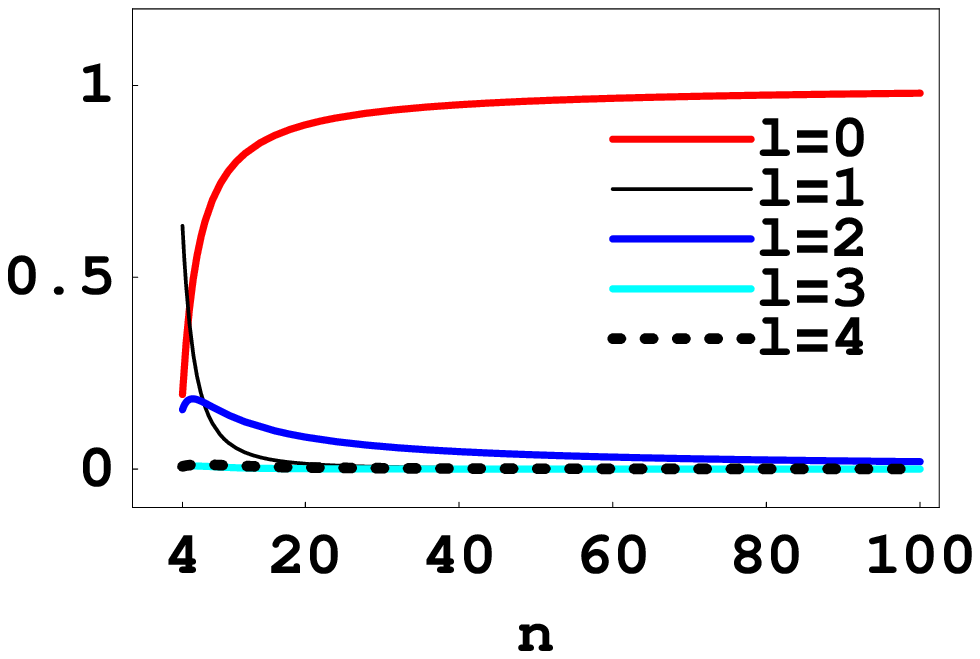}\includegraphics[width=8cm]{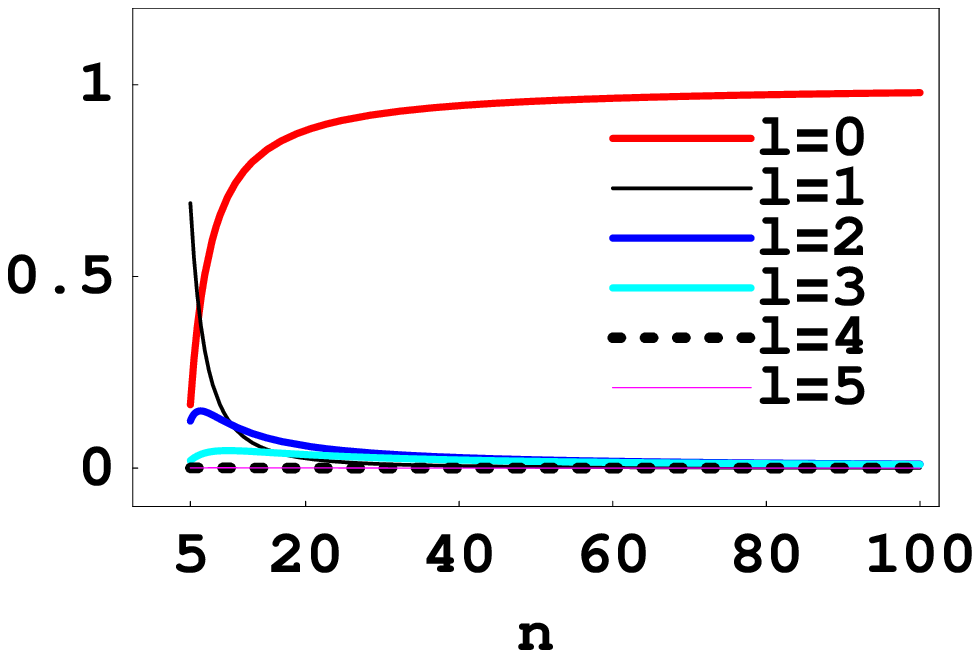}
\caption{ Occupation probabilities $\phi_{l}(n)$ with
$l=0,1,\cdots,L$ as functions of $n$ along the yrast line up to
$L=5$, where the upper left panel is for $L=2$, the upper right one
is for $L=3$, the lower left one is for $L=4$, and the lower right
one is for $L=5$, respectively.}
\end{figure}

\section{Discussion}

In conclusion, entanglement of $M$-body and $M$-body $N$-particle
pure states is defined. The simple entanglement measure formerly
defined for $M=N$ multipartite pure states is extended to the
general cases with any $M$ and $N$. The feature of the measure is
that it only count in entanglement among non-identical particles.
For example, the measure does not count in entanglement among
particles with different internal degrees of freedom on the same
site in lattice models, and so on, which seems to be a natural way
to define entanglement in quantum many-body problems.

Though the measure is defined for pure states, it can easily be
extended to the measure for mixed states $\rho$ by the convex roof
construction with~\cite{[33]}

$$\eta(\rho)=\min_{\Phi}\sum_{i}q_{i}
\eta(\vert\Psi_{i}\rangle\langle\Psi_{i}\vert),\eqno(20)$$ where
$\Phi$ is the set of ensembles realizing the density matrix $\rho$
with $\Phi=( \{q_{i},\vert\Psi_{i}\rangle\}~~\vert~
\rho=\sum_{i}q_{i}
\vert\Psi_{i}\rangle\langle\Psi_{i}\vert$,\\
\noindent $\sum_{i}q_{i}=1)$.

An example of application of the $M$-body $n$-particle pure state
entanglement measure to the Bose-Einstien condensate model with $n$
bosons in a two-dimensional harmonic trap interacting via a fairly
universal contact interaction is provided, which shows that the
measure is effective to quantify entanglement or quantum
correlations among different kinds of particles in the system.
Therefore, the proposed entanglement measure is useful not only in
quantifying entanglement, but also in studying quantum phase
transitions in quantum many-body systems with sufficiently large
number of particles.

\vskip .2cm Support from the U.S. National Science Foundation
(0140300; 0500291), the Southeastern Universitites Research
Association, the Natural Science Foundation of China (10175031;
10575047), and the LSU--LNNU joint research program (C192135) is
acknowledged.

\end{document}